\documentclass[prb,twocolumn,showpacs,showkeys,preprintnumbers,amsmath,amssymb]{revtex4}
\usepackage{graphicx}
\usepackage{dcolumn}
\usepackage{bm}
\usepackage{natbib}
\usepackage[T1]{fontenc}

\begin{document}

\title{Experimental determination of the non-extensive
entropic parameter $q$}

\author{M.S. Reis}
\email{marior@fis.ua.pt}
\author{V.S. Amaral}
\affiliation{Departamento de F\'{i}sica and
CICECO, Universidade de Aveiro, 3810-193 Aveiro,
Portugal}
\author{R.S. Sarthour and I.S. Oliveira}
\affiliation{ Centro Brasileiro de Pesquisas
Físicas, Rua Dr. Xavier Sigaud 150 Urca,
22290-180 Rio de Janeiro-RJ, Brasil}

\date{\today}

\begin{abstract}
We show how to extract the $q$ parameter from
experimental data, considering an inhomogeneous
magnetic system composed by many
Maxwell-Boltzmann homogeneous parts, which after
integration over the whole system recover the
Tsallis non-extensivity. Analyzing the cluster
distribution of La$_{0.7}$Sr$_{0.3}$MnO$_{3}$
manganite, obtained through scanning tunnelling
spectroscopy, we measure the $q$ parameter and
predict the bulk magnetization with good
accuracy. The connection between the Griffiths
phase and non-extensivity is also considered. We
conclude that the entropic parameter embodies
information about the dynamics, the key role to
describe complex systems.
\end{abstract}

\maketitle

It is widely accepted that the statistical
description of a system should be based on its
dynamics; however, it is an information that
indeed does not lie in the Boltzmann entropy.
This fact opens a path for new and different
statistics other than the Boltzmann one. In this
direction, Tsallis thermostatistics has been
strongly used in a number of different
contexts\cite{website}. This framework is
applicable to systems which, broadly speaking,
present at least one of the following properties:
(i) long-range interactions, (ii) long-time
memory, (iii) fractality and (iv) intrinsic
inhomogeneity \cite{website,livro_tsallis}.
Manganese oxides, or simply manganites, seem to
embody three out of these four ingredients: they
present Coulomb long-range interactions
\cite{PRB_64_2001_235127,PRB_64_2001_235128,science_283_1999_2034},
clusters with fractal shapes
\cite{PR_344_2001_1,PRB_66_2002_174436} and
intrinsic inhomogeneity
\cite{PR_344_2001_1,livro_dagotto,PRL_89_2002_237203,PRL_92_2004_126602,nature_428_2004_401}.
Indeed, a sequence of previous publications
\cite{EL_58_2002_42,PRB_66_2002_134417,prb_68_2003_014404},
has shown that the magnetic properties of
manganites can be properly described within a
mean-field approximation using Tsallis
statistics.

In the present Letter, through an analogy to the
works of Beck \cite{PRL_87_2001_180601} and Beck
and Cohen \cite{PA_322_2003_267}, we consider an
inhomogeneous magnetic system composed by many
homogeneous parts with different sizes, each one
of them described by the Maxwell-Boltzmann
statistics. By averaging the magnetization over
the whole system, we recover the Tsallis
non-extensivity. From this point of view, an
analytical relationship between the $q$ parameter
and the moments of the distribution was obtained.
The robustness of the present model was tested
using Scanning Tunnelling Spectroscopy (STS)
conductance maps, where the $q$ parameter could
be obtained and, consequently, the bulk
magnetization predicted. In order to strengthen
the connections between inhomogeneity and
non-extensivity, it is shown that the description
of manganites using Griffiths phase
\cite{PRL_88_2002_197203,prb_68_2003_014411} is
related to the non-extensive treatment, which
contains the dynamics of the system.

We start considering a magnetic system formed by
small regions, or clusters of Maxwell-Boltzmann
bits, each of them with magnetization
$\mathcal{M}$ given by the simple Langevin
function\cite{livro_APG}. The clusters are
distributed in size, and therefore in their net
magnetic moment. Thus, let $f(\mu)$ be the
distribution of magnetic moment of the clusters.
The average magnetization of the sample will be
given by:
\begin{equation}\label{mag_media_momento}
\langle\mathcal{M}\rangle=\int_0^\infty
\mathcal{M}f(\mu)
 d\mu
\end{equation}

Our goal is to connect the above expression to
the non-extensive magnetization, that is
calculated in Ref.
\cite{prb_68_2003_014404,comentario1}:
\begin{equation}\label{mag_generalizada}
\mathcal{M}_q=\frac{\mu_{ne}}{(2-q)}\left[\coth_qx-\frac{1}{x}\right]
\end{equation}
where $x=\mu_{ne}H/kT$, $\coth_q$ is the
generalized $q$-hyperbolic
co-tangent\cite{JPAMG_31_1998_5281}, $q\in\Re$ is
the Tsallis entropic parameter and $\mu_{ne}$
means the magnetic moment of each non-extensive
cluster. The non-extensive correlations lie
inside each cluster, whereas the interactions
inter-clusters remain extensive. This guarantees
that the total magnetization will be additive
\cite{prb_68_2003_014404}.

\emph{Microscopic analysis}: From
Eqs.\ref{mag_media_momento} and
\ref{mag_generalizada}, the average and
non-extensive magnetic susceptibilities can be
derived:
$\langle\chi\rangle=\langle\mu^2\rangle/3kT$ and
$\chi_q=q\mu_{ne}^2/3kT$, as well as the
saturation values of the magnetization:
$\langle\mathcal{M}\rangle_{sat}=\langle\mu\rangle$
and $\mathcal{M}_{q,sat}=\mu_{ne}/(2-q)$. Thus,
equating those limits
($\langle\chi\rangle=\chi_q$ and
$\langle\mathcal{M}\rangle_{sat}=\mathcal{M}_{q,sat}$),
we find a microscopic analytical expression to
the $q$ parameter, in the sense that it is
related to a microscopic information
(distribution of magnetic moments):
\begin{equation}\label{q_momento}
q(2-q)^2=\frac{\langle\mu^2\rangle}{\langle\mu\rangle^2}
\end{equation}
where $\langle\mu\rangle$ and
$\langle\mu^2\rangle$ are the first and second
moments of the distribution $f(\mu)$,
respectively. This result is valid for any
$f(\mu)$, and is analogous to that obtained by
Beck \cite{PRL_87_2001_180601} and Beck and Cohen
\cite{PA_322_2003_267} in other contexts.

\emph{Macroscopic analysis}: From Eq.
\ref{mag_generalizada} we can obtain a
macroscopic analytical expression for the $q$
parameter (similarly to what was done in
\cite{prb_68_2003_014404}), in the sense that it
is now related to macroscopic quantities:
\begin{equation}\label{q_macro}
q(2-q)^2=\frac{3kT\chi}{\mathcal{M}_{sat}^2}
\end{equation}
where $\chi$ and $\mathcal{M}_{sat}$ are the
experimental magnetic susceptibility and
saturation value, respectively. Note that for an
homogenous superparamagnetic-like system, $q=$1.

\emph{Experimental connection with the
microscopic analysis}: The colossal
magnetoresistance (CMR)
effect\cite{livro_dagotto}, usually observed on
manganites, has been explained in terms of
intrinsic
inhomogeneities\cite{livro_dagotto,PR_344_2001_1,PRL_89_2002_237203},
which lead to the formation of insulating and
conducting domains within a single sample, i.e.,
electronic phase separation in a chemically
homogeneous sample. The inhomogeneities alter the
local electronic and magnetic properties of the
sample and should therefore be visible via STS
\cite{nature_416_2002_518,science_285_1999_1540,PRL_89_2002_237203}
or Magnetic Force Microscopy (MFM)
\cite{science_276_1997_2006,science_298_2002_805}.

\begin{figure}
\begin{center}
\includegraphics[width=7cm]{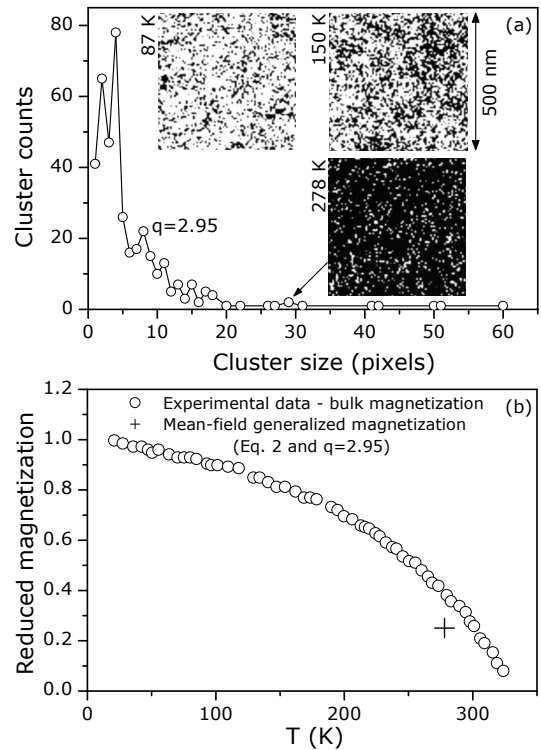}
\end{center}
\caption{(a) Scanning tunnelling spectroscopy
images on La$_{0.7}$Sr$_{0.3}$MnO$_3$/MgO
manganite thin film, after Becker \emph{et
al.}\cite{PRL_89_2002_237203}. White regions mean
conducting (ferromagnetic) clusters and the black
regions stand for insulating (paramagnetic)
phase. The main graphic is the cluster size
distribution of the image at 278 K, proportional
to the cluster magnetic moment distribution,
where, using Eq.\ref{relacao.mi.fi}, $q=2.95$
could be directly obtained. (b) Using the
mean-field (reduced) generalized magnetization
(Eq.\ref{mag_generalizada}: $+$ symbol), we could
predict the bulk magnetization, in a satisfactory
agreement with the measured one ($\circ$ symbol),
in a reduced scale $\mathcal{M}/\mathcal{M}$(18
K)\cite{PRL_89_2002_237203}.} \label{figure1}
\end{figure}

Becker and co-workers\cite{PRL_89_2002_237203}
measured STS in a La$_{0.7}$Sr$_{0.3}$MnO$_3$/MgO
thin film and visualized a domain structure of
conducting (ferromagnetic) and insulating
(paramagnetic) regions with nanometric size. This
manganite has a transition from a metallic phase
(below T$_C$) to an insulating phase (above
T$_C$), with a strong phase
coexistence/competition around T$_C\sim$ 330 K.
The STS conductance maps obtained by those
authors at 87 K, 150 K and 278 K are reproduced
in figure \ref{figure1}(a). From these 1-bit
images (black regions mean
insulating/paramagentic phase and white regions
stand for conducting/ferromagnetic phase), we
have determined the distribution of clusters
size. Considering that the cluster size $\phi$,
measured in \emph{pixels}, is proportional to the
magnetic moment $\mu$ of the cluster, Eq.
\ref{q_momento} can be re-written as:
\begin{equation}\label{relacao.mi.fi}
    \frac{\langle\phi^2\rangle}{\langle\phi\rangle^2}=\frac{\langle\mu^2\rangle}{\langle\mu\rangle^2}=q(2-q)^2
\end{equation}
The conductance map at 278 K has a distribution
of clusters as presented in figure
\ref{figure1}(a), and, using Eq.
\ref{relacao.mi.fi} we obtain from the data
$q=$2.95.

With this value of $q$, the total magnetization
of the system can be predicted, by considering
the mean-field approximation into the generalized
magnetization (Eq.\ref{mag_generalizada}), where
$x=3m_q/t$, $m_q=\mathcal{M}_q/\mu_{ne}$,
$t=T/T_C^{(1)}$ and $T_C^{(1)}=298
K$\cite{EL_58_2002_42} (see refs.
\cite{prb_68_2003_014404,EL_58_2002_42} for
details concerning the mean-field approximation
applied to the non-extensive magnetization). This
procedure results in a satisfactory agreement
between the predicted reduced magnetization
($m_q=$0.25) and the experimental one, obtained
measuring the bulk
magnetization\cite{PRL_89_2002_237203}, as
presented in figure \ref{figure1}(b). The images
at 87 K and 150 K were not analyzed, since the
clusters have already percolated.

The procedure above described shows how to
extract the $q$ parameter from experimental data,
and then how to apply the obtained $q$ parameter
to predict macroscopic quantities of the system.
In addition, these results exemplify the relation
between non-extensivity and microscopic
inhomogeneities. Finally, it is important to
stress that $q$ is related to the dynamics of the
system, since it measures the distribution of
magnetic moments, that contains the dynamics.

\emph{Experimental connections with the
macroscopic analysis}:
Pr$_{0.05}$Ca$_{0.95}$MnO$_3$ ($T_C=$110 K),
LaMnO$_3$ ($T_C=$138 K) and
La$_{0.7}$Ca$_{0.3}$MnO$_3$ ($T_C=$225 K) are
interesting examples of manganites. The inverse
susceptibility as a function of temperature
presents a strong downturn around $T_C$. These
samples were prepared and measured as reported in
Refs\cite{prb_71_2005_144413,JMMM_272_2004_E1671},
except the La-Ca manganite, where the
experimental susceptibility was reproduced from
Ref\cite{prb_68_2003_014411}.

\begin{figure}
\begin{center}
\includegraphics[width=9cm]{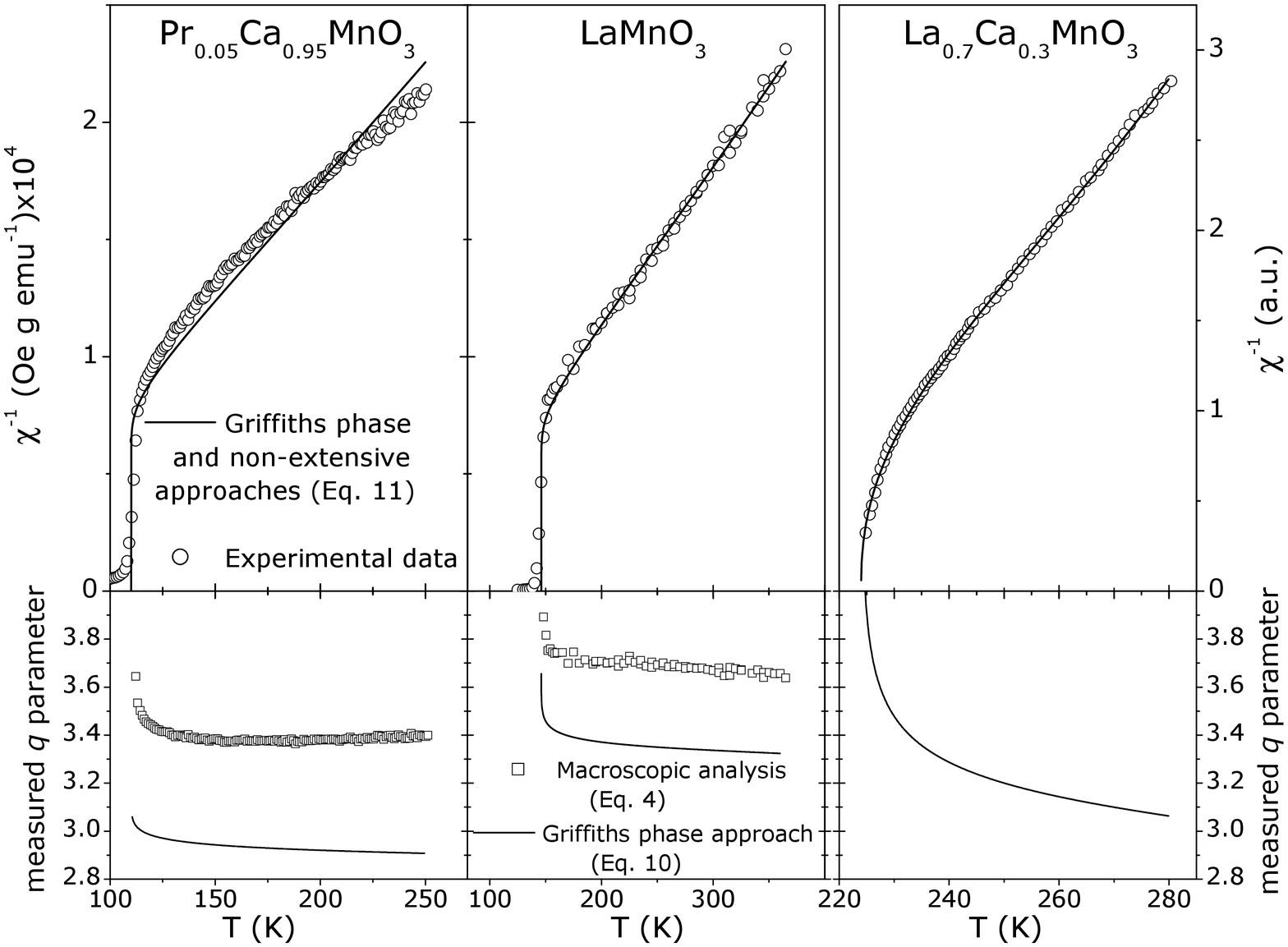}
\end{center}
\caption{Upper panels: Experimental (open
circles) and theoretical (solid line-Eq.
\ref{sus_generalizada}) temperature dependence of
the inverse susceptibility for the
La$_{0.7}$Ca$_{0.3}$MnO$_3$ (after Salamon
\emph{et al.}\cite{prb_68_2003_014411}),
LaMnO$_3$ and Pr$_{0.05}$Ca$_{0.95}$MnO$_3$
manganites. Lower panels: measured $q$ parameter
from the macroscopic analysis and Griffiths phase
approach.} \label{figure2}
\end{figure}

The lower panels of figure \ref{figure2}
($\square$ symbols) present the $q$ parameter
obtained from the macroscopic analysis; using Eq.
\ref{q_macro}, the measured magnetic
susceptibility (experimental data presented in
the upper panels of figure \ref{figure2}) and the
magnetic saturation value for the above cited
manganites (Pr-manganite: 8.9 emu/g and
La-manganite: 8.5 emu/g). This procedure was not
done for the La-Ca manganite, since
Ref.\cite{prb_68_2003_014411} does not provide
the corresponding absolute values for the
magnetic susceptibility.

\begin{figure}
\begin{center}
\includegraphics[width=8cm]{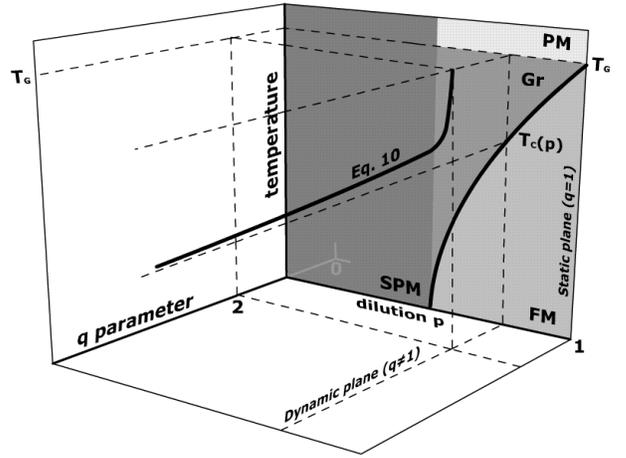}
\end{center}
\caption{Connections between Griffiths phase and
non-extensivity, in a $q-p-T$ space. $p-T$ plane:
usual magnetic phase diagram for a site/bond
diluted magnetic system, where:
SPM-superparamagnetic-like phase, PM-paramagnetic
phase, FM-ferromagnetic phase and Gr-Griffiths
phase. This plane contains the critical
temperature $T_C$ and dilution $p$, basic (and
static) information about the system. $q-T$ plane
contains information related to the dynamics of
the system, through Eq. \ref{razao_medias}. See
text for details.} \label{figure3}
\end{figure}

\emph{Connections with Griffiths phase}: An
extensive number of papers\cite[and references
therein]{pre_60_1999_3823,prb_20_1979_2142,prb_38_1988_11461,PRL_59_1987_586,prl_60_1988_720}
deal with the dynamics of random magnetic
systems, with special attention for a diluted
Ising ferromagnet. The systems of interest are
obtained by starting with an ordinary Ising
model, which contains spins located on the
vertices of a regular lattice. For a bond
dilution, the nearest-neighbor interactions
$J_{ij}$ are independent random variables taking
the values $J$ and 0 with probabilities $p$ and
$1-p$, respectively. For a site dilution,
$J_{ij}=Jc_ic_j$, where $c_{i,j}=1$ or 0, with
probabilities $p$ and $1-p$, respectively. This
diluted ferromagnet is in the Griffiths
phase\cite{prl_60_1988_720,PRL_59_1987_586,prb_38_1988_11461,PRL_88_2002_197203,prb_68_2003_014411}
if its temperature is between the critical
temperature $T_C(p)$ and the critical temperature
$T_G=T_C(1)$ of a pure/non-diluted system. In
figure \ref{figure3}, the dilution
$p$-temperature $T$ plane sketches this scenario.

Salamon and co-workers
\cite{PRL_88_2002_197203,prb_68_2003_014411} have
used the idea of Griffiths singularity to study
manganites, considering a distribution
of the inverse magnetic susceptibility $\lambda$:
\begin{equation}\label{distribuicao_griffiths}
f(\lambda)=\frac{A^{c-1}\lambda^{-c}\exp{(-A/\lambda)}}{\Gamma{(c-1,A/T)}}
\end{equation}
to explain the sharp downturn in the
$\langle\chi\rangle^{-1}$(T) curve (behavior
usually observed in manganites, as displayed in
the upper panels of figure \ref{figure2}). In the
expression above, $c$ is a parameter of the
distribution, $\Gamma$ stands for the Incomplete
Gamma Function and
\begin{equation}\label{a}
A=a\frac{\left(\frac{T}{T_C}-1\right)^{2(1-\beta)}}{\left(1-\frac{T}{T_G}\right)^{2\beta}}
\end{equation}
where $a$ is a free parameter, $\beta=$0.38 is a
critical exponent for the pure system, assumed to
be a 3D Heisenberg-like, and $T_G$ is the
Griffiths temperature
\cite{PRL_88_2002_197203,prb_68_2003_014411,prl_60_1988_720,PRL_59_1987_586,prb_38_1988_11461}.
From Eq. \ref{distribuicao_griffiths} one can
find the inverse average
susceptibility\cite{comentario2}:
\begin{equation}\label{sus_media_final_griffiths}
\langle\chi\rangle^{-1}=A\frac{\Gamma(c-1,A/T)}{\Gamma(c,A/T)}
\end{equation}
that fits the strong downturn usually found on manganites.

The Curie Law $\chi=\mu^2/3kT$ of a small
Maxwell-Boltzmann region tells us that
inhomogeneities in $\chi$ can arise from
distributions of either $\mu$ or $T$ (or both).
Thus, we can obtain from
Eq.\ref{distribuicao_griffiths} a corresponding
distribution of magnetic moments:
\begin{equation}\label{distribuicao_griffiths_momento}
f(\mu)=\left(\frac{A}{3T}\right)^{c-1}\frac{2\;\mu^{2c-3}}{\Gamma(c-1,A/T)}
\exp\left(-\frac{A\mu^2}{3T}\right)
\end{equation}
and, consequently, the $q$ parameter:
\begin{equation}\label{razao_medias}
\frac{\langle\mu^2\rangle}{\langle\mu\rangle^2}=\frac{\Gamma(c,A/T)\;\Gamma(c-1,A/T)}
{\Gamma(c-1/2,A/T)^2}=q(2-q)^2
\end{equation}
This result shows that, since $\langle\mu\rangle$
and $\langle\mu^2\rangle$ are both temperature
dependent, one can expect that $q$ will also be
$T$-dependent. This is expected, since the
distribution of magnetic moments changes as a
function of temperature, changing also the $q$
parameter. Indeed, it can be observed in the STS
images presented in figure \ref{figure1}(a),
reinforcing the idea that $q$ is related to the
dynamics of the system.

From the above, one can write the non-extensive
magnetic susceptibility:
\begin{equation}\label{sus_generalizada}
\chi_q=\frac{q\mu_{ne}^2}{3kT}=\frac{q(2-q)^2\langle\mu\rangle^2}{3kT}=
\frac{\langle\mu^2\rangle}{3kT}=\langle\chi\rangle
\end{equation}
that is equal to the average one. It is important
to note that, once we know $f(\mu)$, the value of
$q$ can be directly obtained, and, consequently,
$\chi_q$. The present approach, as well as those
presented previously, \emph{does not} consider
the entropic index $q$ as a fitting parameter,
but as a known quantity, previously determined
and direct related to the inhomogeneity and
dynamics of the system.

The model above described (Eq.
\ref{sus_generalizada}) is presented in the upper
panels of figure \ref{figure2}. To obtain those
results, the distribution of magnetic moments
$f(\mu)$ presented in Eq.
\ref{distribuicao_griffiths_momento} has the
following parameters: $c=$-0.04, $a=$0.002 K and
$T_G=$510 K, for Pr$_{0.05}$Ca$_{0.95}$MnO$_3$;
$c=$0.01, $a=$4.7x10$^{-8}$ K and $T_G=$555 K,
for LaMnO$_3$; and $c=$0.32, $a=$0.066 K and
$T_G=$335 K, for La$_{0.7}$Ca$_{0.3}$MnO$_3$.
These parameters for the La-Ca manganite are
considerable different from those presented in
Ref.\cite{prb_68_2003_014411}, due to the
remarkable difference between the fit presented
by those authors and that one sketched in figure
\ref{figure2}. The entropic index $q$ (lower
panels of figure \ref{figure2}) \emph{is not} a
free parameter and could be obtained \emph{a
priori}, using Eqs. \ref{q_momento} and
\ref{distribuicao_griffiths_momento}. It is
important to stress the agreement between the
\emph{measured $q$ parameter} using the
macroscopic analysis, and the present approach.

The connection between Griffiths phase and
non-extensivity can be visualized considering the
$q$ parameter as another dimension; an
independent variable in a $q-p-T$ space, as
sketched in figure \ref{figure3}. Analogously to
the Curie temperature, a dilution $p$
characterizes a diluted magnetic system, since
these (static) quantities are unique for a
certain system. If we know only $p$ and $T_C(p)$,
it is not possible to achieve the dynamics of
this diluted magnetic system. For instance, given
a dilution $p$, and consequently a Curie
temperature $T_C(p)$, there are $\binom{N}{pN}$
different ways to dispose these spins,
considering a site dilution problem with $N$
available positions in the lattice. Different
distributions of spins imply different dynamics
and, consequently, different macroscopic
quantities, like magnetization. In this sense,
these quantities ($p$ and $T_C(p)$) are static.
On the other hand, the results presented in this
Letter shown how the $q$ parameter is related to
the dynamics of the system and, once known $q$,
the dynamics is consequently determined. Thus, we
propose that the $p-T$ plane at $q=1$ is a static
plane, whereas the $q-T$ plane (Eq.
\ref{razao_medias}), for a certain value of $p$
(or $T_C(p)$) takes account the dynamics of the
system.
%

Summarizing, in the present Letter we shown that
the $q$ parameter measures the inhomogeneity and
dynamics of a given inhomogeneous magnetic
system. From the measured $q$, obtained from
scanning tunnelling spectroscopy on manganites (a
microscopic information), one can predict a
thermodynamic quantity, the bulk magnetization (a
macroscopic information); the entropic parameter
contains the dynamics, connecting the microscopic
and macroscopic information. The present model
was also successfully applied to other manganites
(bulk and thin films) and melt-spun granular
alloys, reinforcing the conclusions made.

We acknowledge CAPES-Brasil, CNPq-Brasil and
GRICES-Portugal. We are also thankful to P.B.
Tavares, A.M.L. Lopes, M.P. Albuquerque and A.R.
Gesualdi for valuable discussions.


\begin{thebibliography}{33}
\expandafter\ifx\csname
natexlab\endcsname\relax\def\natexlab#1{#1}\fi
\expandafter\ifx\csname
bibnamefont\endcsname\relax
  \def\bibnamefont#1{#1}\fi
\expandafter\ifx\csname
bibfnamefont\endcsname\relax
  \def\bibfnamefont#1{#1}\fi
\expandafter\ifx\csname
citenamefont\endcsname\relax
  \def\citenamefont#1{#1}\fi
\expandafter\ifx\csname url\endcsname\relax
  \def\url#1{\texttt{#1}}\fi
\expandafter\ifx\csname
urlprefix\endcsname\relax\def\urlprefix{URL }\fi
\providecommand{\bibinfo}[2]{#2}
\providecommand{\eprint}[2][]{\url{#2}}

\bibitem[{web()}]{website}
\bibinfo{note}{For a complete and updated list of references, see the web site:
  \texttt{tsallis.cat.cbpf.br/biblio.htm}}.

\bibitem[{\citenamefont{Tsallis}(2001)}]{livro_tsallis}
\bibinfo{author}{\bibfnamefont{C.}~\bibnamefont{Tsallis}},
  \emph{\bibinfo{title}{Nonextensive Statistical Mechanics and Its
  Applications}} (\bibinfo{publisher}{Springer-Verlag},
  \bibinfo{address}{Heidelberg}, \bibinfo{year}{2001}), chap.
  \bibinfo{chapter}{Nonextensive Statistical Mechanics and Thermodynamics:
  Historical Background and Present Status}, eds. S.~Abe and Y.~Okamoto.

\bibitem[{\citenamefont{Lorenzana
  et~al.}(2001{\natexlab{a}})\citenamefont{Lorenzana, Castellani, and
  Castro}}]{PRB_64_2001_235127}
\bibinfo{author}{\bibfnamefont{J.}~\bibnamefont{Lorenzana}},
  \bibinfo{author}{\bibfnamefont{C.}~\bibnamefont{Castellani}},
  \bibnamefont{and} \bibinfo{author}{\bibfnamefont{C.~D.}
  \bibnamefont{Castro}}, \bibinfo{journal}{Phys. Rev. B}
  \textbf{\bibinfo{volume}{64}}, \bibinfo{pages}{235127}
  (\bibinfo{year}{2001}{\natexlab{a}}).

\bibitem[{\citenamefont{Lorenzana
  et~al.}(2001{\natexlab{b}})\citenamefont{Lorenzana, Castellani, and
  Castro}}]{PRB_64_2001_235128}
\bibinfo{author}{\bibfnamefont{J.}~\bibnamefont{Lorenzana}},
  \bibinfo{author}{\bibfnamefont{C.}~\bibnamefont{Castellani}},
  \bibnamefont{and} \bibinfo{author}{\bibfnamefont{C.~D.}
  \bibnamefont{Castro}}, \bibinfo{journal}{Phys. Rev. B}
  \textbf{\bibinfo{volume}{64}}, \bibinfo{pages}{235128}
  (\bibinfo{year}{2001}{\natexlab{b}}).

\bibitem[{\citenamefont{Moreo et~al.}(1999)\citenamefont{Moreo, Yunoki, and
  Dagotto}}]{science_283_1999_2034}
\bibinfo{author}{\bibfnamefont{A.}~\bibnamefont{Moreo}},
  \bibinfo{author}{\bibfnamefont{S.}~\bibnamefont{Yunoki}}, \bibnamefont{and}
  \bibinfo{author}{\bibfnamefont{E.}~\bibnamefont{Dagotto}},
  \bibinfo{journal}{Science} \textbf{\bibinfo{volume}{283}},
  \bibinfo{pages}{2034} (\bibinfo{year}{1999}).

\bibitem[{\citenamefont{Dagotto et~al.}(2001)\citenamefont{Dagotto, Hotta, and
  Moreo}}]{PR_344_2001_1}
\bibinfo{author}{\bibfnamefont{E.}~\bibnamefont{Dagotto}},
  \bibinfo{author}{\bibfnamefont{T.}~\bibnamefont{Hotta}}, \bibnamefont{and}
  \bibinfo{author}{\bibfnamefont{A.}~\bibnamefont{Moreo}},
  \bibinfo{journal}{Phys. Rep.} \textbf{\bibinfo{volume}{344}},
  \bibinfo{pages}{1} (\bibinfo{year}{2001}).

\bibitem[{\citenamefont{Ausloos et~al.}(2002)\citenamefont{Ausloos, Hubert,
  Dorbolo, Gilabert, and Cloots}}]{PRB_66_2002_174436}
\bibinfo{author}{\bibfnamefont{M.}~\bibnamefont{Ausloos}},
  \bibinfo{author}{\bibfnamefont{L.}~\bibnamefont{Hubert}},
  \bibinfo{author}{\bibfnamefont{S.}~\bibnamefont{Dorbolo}},
  \bibinfo{author}{\bibfnamefont{A.}~\bibnamefont{Gilabert}}, \bibnamefont{and}
  \bibinfo{author}{\bibfnamefont{R.}~\bibnamefont{Cloots}},
  \bibinfo{journal}{Phys. Rev. B} \textbf{\bibinfo{volume}{66}},
  \bibinfo{pages}{174436} (\bibinfo{year}{2002}).

\bibitem[{\citenamefont{Dagotto}(2003)}]{livro_dagotto}
\bibinfo{author}{\bibfnamefont{E.}~\bibnamefont{Dagotto}},
  \emph{\bibinfo{title}{Nanoscale phase separation and colossal
  magnetoresistance: The physics of manganites and related compounds.}}
  (\bibinfo{publisher}{Springer-Verlag}, \bibinfo{address}{Heidelberg},
  \bibinfo{year}{2003}).

\bibitem[{\citenamefont{Becker et~al.}(2002)\citenamefont{Becker, Streng, Luo,
  Moshnyaga, Damaschke, Shannon, and Samwer}}]{PRL_89_2002_237203}
\bibinfo{author}{\bibfnamefont{T.}~\bibnamefont{Becker}},
  \bibinfo{author}{\bibfnamefont{C.}~\bibnamefont{Streng}},
  \bibinfo{author}{\bibfnamefont{Y.}~\bibnamefont{Luo}},
  \bibinfo{author}{\bibfnamefont{V.}~\bibnamefont{Moshnyaga}},
  \bibinfo{author}{\bibfnamefont{B.}~\bibnamefont{Damaschke}},
  \bibinfo{author}{\bibfnamefont{N.}~\bibnamefont{Shannon}}, \bibnamefont{and}
  \bibinfo{author}{\bibfnamefont{K.}~\bibnamefont{Samwer}},
  \bibinfo{journal}{Phys. Rev. Lett.} \textbf{\bibinfo{volume}{89}},
  \bibinfo{pages}{237203} (\bibinfo{year}{2002}).

\bibitem[{\citenamefont{Kumar and Majumdar}(2004)}]{PRL_92_2004_126602}
\bibinfo{author}{\bibfnamefont{S.}~\bibnamefont{Kumar}} \bibnamefont{and}
  \bibinfo{author}{\bibfnamefont{P.}~\bibnamefont{Majumdar}},
  \bibinfo{journal}{Phys. Rev. Lett.} \textbf{\bibinfo{volume}{92}},
  \bibinfo{pages}{126602} (\bibinfo{year}{2004}).

\bibitem[{\citenamefont{Ahn et~al.}(2004)\citenamefont{Ahn, Lookman, and
  Bishop}}]{nature_428_2004_401}
\bibinfo{author}{\bibfnamefont{K.~H.} \bibnamefont{Ahn}},
  \bibinfo{author}{\bibfnamefont{T.}~\bibnamefont{Lookman}}, \bibnamefont{and}
  \bibinfo{author}{\bibfnamefont{A.~R.} \bibnamefont{Bishop}},
  \bibinfo{journal}{Nature} \textbf{\bibinfo{volume}{428}},
  \bibinfo{pages}{401} (\bibinfo{year}{2004}).

\bibitem[{\citenamefont{Reis et~al.}(2002{\natexlab{a}})\citenamefont{Reis,
  Freitas, Orlando, Lenzi, and Oliveira}}]{EL_58_2002_42}
\bibinfo{author}{\bibfnamefont{M.~S.} \bibnamefont{Reis}},
  \bibinfo{author}{\bibfnamefont{J.~C.~C.} \bibnamefont{Freitas}},
  \bibinfo{author}{\bibfnamefont{M.~T.~D.} \bibnamefont{Orlando}},
  \bibinfo{author}{\bibfnamefont{E.~K.} \bibnamefont{Lenzi}}, \bibnamefont{and}
  \bibinfo{author}{\bibfnamefont{I.~S.} \bibnamefont{Oliveira}},
  \bibinfo{journal}{Europhys. Lett.} \textbf{\bibinfo{volume}{58}},
  \bibinfo{pages}{42} (\bibinfo{year}{2002}{\natexlab{a}}).

\bibitem[{\citenamefont{Reis et~al.}(2002{\natexlab{b}})\citenamefont{Reis,
  Ara\'{u}jo, Amaral, Lenzi, and Oliveira}}]{PRB_66_2002_134417}
\bibinfo{author}{\bibfnamefont{M.~S.} \bibnamefont{Reis}},
  \bibinfo{author}{\bibfnamefont{J.~P.} \bibnamefont{Ara\'{u}jo}},
  \bibinfo{author}{\bibfnamefont{V.~S.} \bibnamefont{Amaral}},
  \bibinfo{author}{\bibfnamefont{E.~K.} \bibnamefont{Lenzi}}, \bibnamefont{and}
  \bibinfo{author}{\bibfnamefont{I.~S.} \bibnamefont{Oliveira}},
  \bibinfo{journal}{Phys. Rev. B} \textbf{\bibinfo{volume}{66}},
  \bibinfo{pages}{134417} (\bibinfo{year}{2002}{\natexlab{b}}).

\bibitem[{\citenamefont{Reis et~al.}(2003)\citenamefont{Reis, Amaral,
  Ara\'{u}jo, and Oliveira}}]{prb_68_2003_014404}
\bibinfo{author}{\bibfnamefont{M.~S.} \bibnamefont{Reis}},
  \bibinfo{author}{\bibfnamefont{V.~S.} \bibnamefont{Amaral}},
  \bibinfo{author}{\bibfnamefont{J.~P.} \bibnamefont{Ara\'{u}jo}},
  \bibnamefont{and} \bibinfo{author}{\bibfnamefont{I.~S.}
  \bibnamefont{Oliveira}}, \bibinfo{journal}{Phys. Rev. B}
  \textbf{\bibinfo{volume}{68}}, \bibinfo{pages}{014404}
  (\bibinfo{year}{2003}).

\bibitem[{\citenamefont{Beck}(2001)}]{PRL_87_2001_180601}
\bibinfo{author}{\bibfnamefont{C.}~\bibnamefont{Beck}}, \bibinfo{journal}{Phys.
  Rev. Lett.} \textbf{\bibinfo{volume}{87}}, \bibinfo{pages}{180601}
  (\bibinfo{year}{2001}).

\bibitem[{\citenamefont{Beck and Cohen}(2003)}]{PA_322_2003_267}
\bibinfo{author}{\bibfnamefont{C.}~\bibnamefont{Beck}} \bibnamefont{and}
  \bibinfo{author}{\bibfnamefont{E.~G.~D.} \bibnamefont{Cohen}},
  \bibinfo{journal}{Physica A} \textbf{\bibinfo{volume}{322}},
  \bibinfo{pages}{267} (\bibinfo{year}{2003}).

\bibitem[{\citenamefont{Salamon et~al.}(2002)\citenamefont{Salamon, Lin, and
  Chun}}]{PRL_88_2002_197203}
\bibinfo{author}{\bibfnamefont{M.~B.} \bibnamefont{Salamon}},
  \bibinfo{author}{\bibfnamefont{P.}~\bibnamefont{Lin}}, \bibnamefont{and}
  \bibinfo{author}{\bibfnamefont{S.~H.} \bibnamefont{Chun}},
  \bibinfo{journal}{Phys. Rev. Lett.} \textbf{\bibinfo{volume}{88}},
  \bibinfo{pages}{197203} (\bibinfo{year}{2002}).

\bibitem[{\citenamefont{Salamon and Chun}(2003)}]{prb_68_2003_014411}
\bibinfo{author}{\bibfnamefont{M.}~\bibnamefont{Salamon}} \bibnamefont{and}
  \bibinfo{author}{\bibfnamefont{S.}~\bibnamefont{Chun}},
  \bibinfo{journal}{Phys. Rev. B} \textbf{\bibinfo{volume}{68}},
  \bibinfo{pages}{014411} (\bibinfo{year}{2003}).

\bibitem[{\citenamefont{Guimarães}(1998)}]{livro_APG}
\bibinfo{author}{\bibfnamefont{A.~P.} \bibnamefont{Guimarães}},
  \emph{\bibinfo{title}{Magnetism and Magnetic Resonance in Solids}}
  (\bibinfo{publisher}{John Wiley}, \bibinfo{address}{New York},
  \bibinfo{year}{1998}).

\bibitem[{com({\natexlab{a}})}]{comentario1}
\bibinfo{note}{In our previous work\cite{prb_68_2003_014404}, we derived the
  two-branched Generalized Langevin Function. However, further calculations
  lead us to a more concise and general expression, valid for any real value of
  $q$ and $x$ (Eq.\ref{mag_generalizada})}.

\bibitem[{\citenamefont{Borges}(1998)}]{JPAMG_31_1998_5281}
\bibinfo{author}{\bibfnamefont{E.~P.} \bibnamefont{Borges}},
  \bibinfo{journal}{J. Phys. A: Math. Gen.} \textbf{\bibinfo{volume}{31}},
  \bibinfo{pages}{5281} (\bibinfo{year}{1998}).

\bibitem[{\citenamefont{Renner et~al.}(2002)\citenamefont{Renner, Aeppli, Kim,
  Soh, and Cheong}}]{nature_416_2002_518}
\bibinfo{author}{\bibfnamefont{C.}~\bibnamefont{Renner}},
  \bibinfo{author}{\bibfnamefont{G.}~\bibnamefont{Aeppli}},
  \bibinfo{author}{\bibfnamefont{B.}~\bibnamefont{Kim}},
  \bibinfo{author}{\bibfnamefont{Y.}~\bibnamefont{Soh}}, \bibnamefont{and}
  \bibinfo{author}{\bibfnamefont{S.}~\bibnamefont{Cheong}},
  \bibinfo{journal}{Nature} \textbf{\bibinfo{volume}{416}},
  \bibinfo{pages}{518} (\bibinfo{year}{2002}).

\bibitem[{\citenamefont{Fath et~al.}(1999)\citenamefont{Fath, Freisem,
  Menovsky, Tomioka, Aarts, and Mydosh}}]{science_285_1999_1540}
\bibinfo{author}{\bibfnamefont{M.}~\bibnamefont{Fath}},
  \bibinfo{author}{\bibfnamefont{S.}~\bibnamefont{Freisem}},
  \bibinfo{author}{\bibfnamefont{A.~A.} \bibnamefont{Menovsky}},
  \bibinfo{author}{\bibfnamefont{Y.}~\bibnamefont{Tomioka}},
  \bibinfo{author}{\bibfnamefont{J.}~\bibnamefont{Aarts}}, \bibnamefont{and}
  \bibinfo{author}{\bibfnamefont{J.~A.} \bibnamefont{Mydosh}},
  \bibinfo{journal}{Science} \textbf{\bibinfo{volume}{285}},
  \bibinfo{pages}{1540} (\bibinfo{year}{1999}).

\bibitem[{\citenamefont{Lu et~al.}(1997)\citenamefont{Lu, Chen, and
  deLozanne}}]{science_276_1997_2006}
\bibinfo{author}{\bibfnamefont{Q.}~\bibnamefont{Lu}},
  \bibinfo{author}{\bibfnamefont{C.}~\bibnamefont{Chen}}, \bibnamefont{and}
  \bibinfo{author}{\bibfnamefont{A.}~\bibnamefont{deLozanne}},
  \bibinfo{journal}{Science} \textbf{\bibinfo{volume}{276}},
  \bibinfo{pages}{2006} (\bibinfo{year}{1997}).

\bibitem[{\citenamefont{Zhang et~al.}(2002)\citenamefont{Zhang, Israel, Biswas,
  Greene, and deLozanne}}]{science_298_2002_805}
\bibinfo{author}{\bibfnamefont{L.}~\bibnamefont{Zhang}},
  \bibinfo{author}{\bibfnamefont{C.}~\bibnamefont{Israel}},
  \bibinfo{author}{\bibfnamefont{A.}~\bibnamefont{Biswas}},
  \bibinfo{author}{\bibfnamefont{R.~L.} \bibnamefont{Greene}},
  \bibnamefont{and}
  \bibinfo{author}{\bibfnamefont{A.}~\bibnamefont{deLozanne}},
  \bibinfo{journal}{Science} \textbf{\bibinfo{volume}{298}},
  \bibinfo{pages}{805} (\bibinfo{year}{2002}).

\bibitem[{\citenamefont{Reis et~al.}(2005)\citenamefont{Reis, Amaral,
  Ara\'{u}jo, Tavares, Gomes, and Oliveira}}]{prb_71_2005_144413}
\bibinfo{author}{\bibfnamefont{M.~S.} \bibnamefont{Reis}},
  \bibinfo{author}{\bibfnamefont{V.~S.} \bibnamefont{Amaral}},
  \bibinfo{author}{\bibfnamefont{J.~P.} \bibnamefont{Ara\'{u}jo}},
  \bibinfo{author}{\bibfnamefont{P.~B.} \bibnamefont{Tavares}},
  \bibinfo{author}{\bibfnamefont{A.~M.} \bibnamefont{Gomes}}, \bibnamefont{and}
  \bibinfo{author}{\bibfnamefont{I.~S.} \bibnamefont{Oliveira}},
  \bibinfo{journal}{Phys. Rev. B} \textbf{\bibinfo{volume}{71}},
  \bibinfo{pages}{144413} (\bibinfo{year}{2005}).

\bibitem[{\citenamefont{Lopes et~al.}(2004)\citenamefont{Lopes, Araujo, Rita,
  Correia, Amaral, Suryanarayanan, and ISOLDE}}]{JMMM_272_2004_E1671}
\bibinfo{author}{\bibfnamefont{A.~M.~L.} \bibnamefont{Lopes}},
  \bibinfo{author}{\bibfnamefont{J.~P.} \bibnamefont{Araujo}},
  \bibinfo{author}{\bibfnamefont{E.}~\bibnamefont{Rita}},
  \bibinfo{author}{\bibfnamefont{J.~G.} \bibnamefont{Correia}},
  \bibinfo{author}{\bibfnamefont{V.~S.} \bibnamefont{Amaral}},
  \bibinfo{author}{\bibfnamefont{R.}~\bibnamefont{Suryanarayanan}},
  \bibnamefont{and} \bibinfo{author}{\bibnamefont{ISOLDE}},
  \bibinfo{journal}{J. Magn. Magn. Mater.} \textbf{\bibinfo{volume}{272}},
  \bibinfo{pages}{E1671} (\bibinfo{year}{2004}).

\bibitem[{\citenamefont{Mazzeo and Kuhn}(1999)}]{pre_60_1999_3823}
\bibinfo{author}{\bibfnamefont{G.}~\bibnamefont{Mazzeo}} \bibnamefont{and}
  \bibinfo{author}{\bibfnamefont{R.}~\bibnamefont{Kuhn}},
  \bibinfo{journal}{Phys. Rev. E} \textbf{\bibinfo{volume}{60}},
  \bibinfo{pages}{3823} (\bibinfo{year}{1999}).

\bibitem[{\citenamefont{Thorpe and McGurn}(1979)}]{prb_20_1979_2142}
\bibinfo{author}{\bibfnamefont{M.}~\bibnamefont{Thorpe}} \bibnamefont{and}
  \bibinfo{author}{\bibfnamefont{A.}~\bibnamefont{McGurn}},
  \bibinfo{journal}{Phys. Rev. B} \textbf{\bibinfo{volume}{20}},
  \bibinfo{pages}{2142} (\bibinfo{year}{1979}).

\bibitem[{\citenamefont{Bray and Rodgers}(1988)}]{prb_38_1988_11461}
\bibinfo{author}{\bibfnamefont{A.}~\bibnamefont{Bray}} \bibnamefont{and}
  \bibinfo{author}{\bibfnamefont{G.}~\bibnamefont{Rodgers}},
  \bibinfo{journal}{Phys. Rev. B} \textbf{\bibinfo{volume}{38}},
  \bibinfo{pages}{11461} (\bibinfo{year}{1988}).

\bibitem[{\citenamefont{Bray}(1987)}]{PRL_59_1987_586}
\bibinfo{author}{\bibfnamefont{A.}~\bibnamefont{Bray}}, \bibinfo{journal}{Phys.
  Rev. Lett.} \textbf{\bibinfo{volume}{59}}, \bibinfo{pages}{586}
  (\bibinfo{year}{1987}).

\bibitem[{\citenamefont{Bray}(1988)}]{prl_60_1988_720}
\bibinfo{author}{\bibfnamefont{A.}~\bibnamefont{Bray}}, \bibinfo{journal}{Phys.
  Rev. Lett.} \textbf{\bibinfo{volume}{60}}, \bibinfo{pages}{720}
  (\bibinfo{year}{1988}).

\bibitem[{com({\natexlab{b}})}]{comentario2}
\bibinfo{note}{In a different fashion as presented by Salamon and co-workers
  \cite{PRL_88_2002_197203,prb_68_2003_014411}, we derived an analytical and
  closed expression to the inverse average susceptibility.}

\end{thebibliography}
\end{document}